\documentclass[runningheads]{svmult}

\usepackage{makeidx}   
\usepackage{graphicx}  
\usepackage{subeqnar}  
\usepackage{multicol}  
\usepackage{physprbb}  
\makeindex             

\begin{document}
%
\title*{Stellar Masses of High--Redshift Galaxies}
\toctitle{Stellar Masses of High--Redshift Galaxies}
%
%
\titlerunning{Stellar Masses of High--Redshift Galaxies}
%
\author{Casey Papovich\inst{1,2}
\and Mark Dickinson\inst{2}
\and Henry C.\ Ferguson\inst{2}
}
\authorrunning{Papovich, Dickinson, \& Ferguson}
%
%
\institute{Steward Observatory, Univ.\ of Arizona, Tucson AZ 85721, USA 
\and Space Telescope Science Inst., Baltimore MD 21218, USA}

\maketitle              

\noindent{\small To appear in \textit{The Mass of Galaxies at Low and High Redshift}, eds.\ R. Bender \& A. Renzini (ESO Astrophysics Symposia, Springer-Verlag), Venice, 24-26 Oct 2001.}

\begin{abstract}
We present constraints on the stellar--mass distribution of distant
galaxies.  These stellar mass estimates derive from fitting
population--synthesis models to the galaxies' observed multi-band
spectrophotometry.  We discuss the complex uncertainties (both
statistical and systematic) that are inherent to this method, and
offer future prospects to improve the constraints.  Typical
uncertainties for galaxies at $z\sim 2.5$ are $\delta(\log \mathcal{M})
\sim 0.3$~dex (statistical), and factors of $\mathrel{\hbox{\rlap{\hbox{\lower4pt\hbox{$\sim$}}}\hbox{$>$}}} 3$ (systematic).
By applying this method to a catalog of NICMOS--selected galaxies in
the Hubble Deep Field North, we generally find a lack of
high--redshift galaxies ($z\mathrel{\hbox{\rlap{\hbox{\lower4pt\hbox{$\sim$}}}\hbox{$>$}}} 2$) with masses comparable to those
of present--day ``$L^\ast$'' galaxies.  At $z\mathrel{\hbox{\rlap{\hbox{\lower4pt\hbox{$\sim$}}}\hbox{$<$}}} 1.8$, galaxies with
$L^\ast$--sized masses do emerge, but with a number--density below
that at the present epoch.  Thus, it seems massive, present--day
galaxies were not fully assembled by $z\sim 2.5$, and that further
star formation and/or merging are required to assemble them from these
high--redshift  progenitors.  Future progress on this subject will
greatly benefit from upcoming surveys such as those planned with
\emph{HST}/ACS and \emph{SIRTF}.
\end{abstract}

\section{Introduction and Motivation}

With current observations and those of the near future, we are able to
observe distant galaxies ($z\mathrel{\hbox{\rlap{\hbox{\lower4pt\hbox{$\sim$}}}\hbox{$>$}}} 2$) in their primeval stages, i.e.,
at an era when they are vigorously assembling their stellar content.
However, no conclusive picture has yet emerged to describe how these
high--redshift galaxies fit into the ancestral history of the
present--day galaxy population.  By measuring the stellar--mass
distribution (which contains a complete historical record of star
formation) for galaxies as a function of redshift, one can directly
probe the global, mass--assembly history.  This provides a stringent
test for cosmological models that recount how high--redshift galaxies
evolve into the present--day galaxy population.

However, a galaxy's stellar mass is \emph{not} a directly measurable
quantity: it must be inferred from models of the galaxy's
mass--to--light ratios and the observed multi-band photometry.  In this
contribution, we discuss the method used to obtain stellar--mass
estimates of distant galaxies and some the underlying caveats inherent
in the process.  We then present results from applying this method to
a NICMOS--selected sample of galaxies in the \emph{Hubble Deep Field
North} (HDF--N).

\section{Methodology}

As the sample for our study, we have investigated the stellar--mass
content of galaxies in the HDF--N  using the multi-band photometry from
the \textit{HST}/WFPC2 (\hbox{$U_{300}$}\hbox{$B_{450}$}\hbox{$V_{606}$}\hbox{$I_{814}$}),  \textit{HST}/NIC3 (\hbox{$J_{110}$}\hbox{$H_{160}$}), and
ground--based $K_s$~\cite{dic01}.  We initially focused on a sample of
31 ``Lyman--break galaxies'' (LBGs) with $2 \mathrel{\hbox{\rlap{\hbox{\lower4pt\hbox{$\sim$}}}\hbox{$<$}}} z \mathrel{\hbox{\rlap{\hbox{\lower4pt\hbox{$\sim$}}}\hbox{$<$}}} 3.5$
~\cite{pap01} and fit their spectrophotometry with a suite of
stellar--population--synthesis models~\cite{bru93,bru01}, varying the
age, SFR ``e--folding'' timescale ($\tau_\mathrm{SF}$), extinction
($A_\lambda$), and stellar mass; and also considered a range of
metallicities ($0.001 - 3$~$Z_\odot$), and IMF (Salpeter; Scalo; Miller \&
Scalo). In general, we found only loose constraints on the parameters
of the galaxies' stellar populations (i.e., age, $\tau_\mathrm{SF}$,
$A_\lambda$, $Z$, and IMF).  However, we derive fairly robust
constraints for galaxy stellar masses (typical \emph{statistical}
uncertainty is $\sim 0.3$~dex).

\begin{figure}
\begin{center}
\leavevmode
\includegraphics[width=0.667\textwidth]{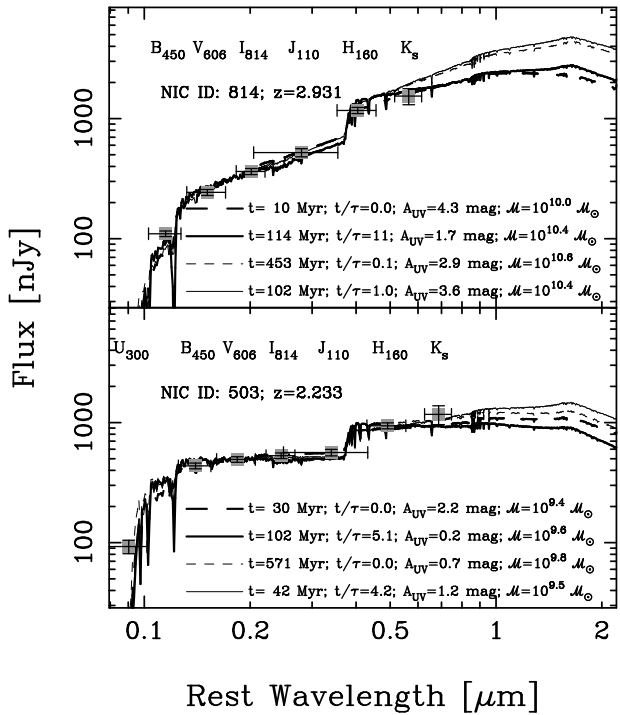}
\end{center}
\caption{Stellar--population--synthesis--model fits to the observed
photometry for two LBGs in the HDF--N.  Each panel shows four models
(with parameters inset), all of which fit the observed photometry at
the 95\% confidence level.  \label{fig:multi}}
\end{figure}

\begin{figure}
\begin{center}
\leavevmode
\includegraphics[width=1\textwidth]{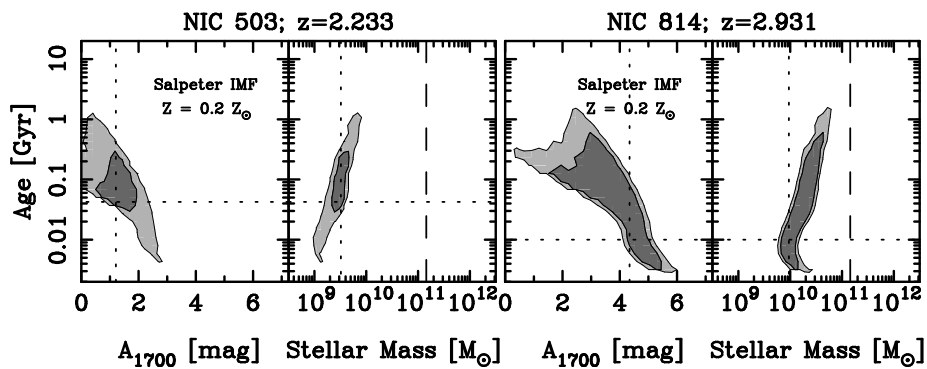}
\end{center}
\caption{Confidence intervals on the fitted parameters for the two
galaxies in fig.~\ref{fig:multi}.  The equivalent 68\% and 95\%
confidence intervals of extinction (at $1700$~\AA) and stellar mass
are plotted versus the population age.  Dotted lines indicate the
best--fit solutions. The dashed line shows the characteristic stellar
mass of a present--day ``$L^\ast$'' galaxy~\cite{col01}.}
\label{fig:contour}
\end{figure}

Figure~\ref{fig:multi} illustrates the range of model parameters
capable of fitting the observed photometry for two of the galaxies in
the sample.  Note that for each galaxy, there exist acceptable model
fits with a wide range (i.e., more than an order of magnitude) of
population age, $\tau_\mathrm{SF}$, and extinction.  However, the
stellar mass fits remain roughly constant [$\delta(\log \mathcal{M})
\sim 0.3$~dex].  This is also depicted in fig.~\ref{fig:contour}.
Note also that while the models fit the observed photometry (out to
rest--frame $\sim 6000$~\AA), they diverge strongly for $\lambda \mathrel{\hbox{\rlap{\hbox{\lower4pt\hbox{$\sim$}}}\hbox{$>$}}}
1$~$\mu$m.  This is generally true for the entire galaxy sample:
there are large degeneracies in model age and extinction, which
translates to a statistical uncertainty on the stellar mass estimates.
Improved constraints would be possible with the incorporation of
independent measurements of the instantaneous star--formation rate
(e.g., nebular emission lines, FIR flux measures, etc.).

Although the statistical uncertainties on the galaxies' stellar--mass
estimates are generally low, there remain inherent systematic
uncertainties that must be considered when interpreting the results.
Some of this arises from assumptions in the population--synthesis
models (e.g., metallicity, IMF; see~\cite{pap01}).  The metallicities
of high--redshift galaxies are only weakly constrained; optical and
near--IR spectra suggest $\sim 1/4 - 1/3$~$Z_\odot$ (e.g., \cite{pet01}).
Varying the metallicity assumed in the synthesis models causes
systematic shifts in the distribution of best--fit--model parameters,
including $\sim 0.3$~dex in the inferred stellar masses.  Similarly,
the IMF at high--redshifts is essentially unconstrained. We find that
a gamut of IMF models (Salpeter; Scalo; or Miller--Scalo) are all
statistically consistent with the data; no one model is more
preferred.  The Scalo and Miller \& Scalo IMFs systematically favor
younger ages, lower extinctions, and somewhat higher stellar--masses.
The lack of knowledge of the low--mass end of the IMF is also
problematic.  E.g., for a Salpeter IMF with a low--mass cutoff of
$1$~$\mathcal{M}_\odot$, the total stellar mass would be 39\% that derived with a
cutoff at $0.1$~$\mathcal{M}_\odot$.

\begin{figure}
\begin{center}
\leavevmode
\includegraphics[width=0.85\textwidth]{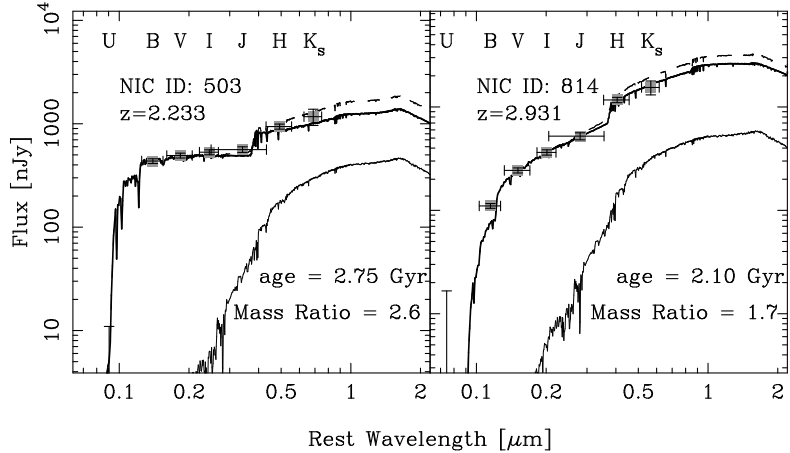}
\end{center}
\caption{Schematic illustration showing the effect on the
stellar--mass estimates when adding a component corresponding to a
maximally old stellar population.  The thick line shows the
best--fitting ``young'' model, and the thin line shows the maximum
allowable contribution from an additional, old stellar population
whose age is that of the universe at the observed redshift.  The mass
ratios of the old--to--young stellar populations are given in the
panels.  The dashed lines show the superposition of the two
models.\label{fig:old}}
\end{figure}

Systematics also arise from the assumptions of the galaxies'
star--formation histories. All results presented thus far have used a
monotonic, exponentially decaying (or constant) star--formation
history.  Such models likely only pertain to the youngest (and most
dominant) stellar populations and as such neglect the contribution
from any underlying, older stellar population.  Because the
single--component star--formation histories pertain to the youngest
stellar populations, they arguably provide a \emph{minimal} inferred
$\mathcal{M}/L$ --- and thus mass ---  for the galaxy.  One can
consider the flux contribution of an old stellar population from
previous star--formation that is hidden ``beneath the glare'' of the
young stars.  We have investigated this effect by considering the sum
of the fluxes from a maximally--old stellar component to that from the
single--component models. The old--stellar component predominantly
contributes to the flux longward of $\sim 6000$~\AA\ (see
fig.~\ref{fig:old}).  These two--component models yield a scenario
where some fraction of the galaxies' stellar populations formed in a
``burst'' in the distant past.  Such a scenario produces a maximal
inferred $\mathcal{M}/L$, and thus translates to an upper bound to the
galaxies' total  stellar mass.  For the HDF--N LBGs, the
two--component models on average provide stellar mass estimates $\sim
3$ times those from the single--component fits. Such a scenario is
somewhat nonphysical (it assumes that most of the galaxies' observed
stellar mass formed at $z\approx \infty$ and has since evolved
passively), and considering this population merely serves as a
fiducial with which to constrain the upper bound on the galaxy stellar
masses.

\section{Discussion and Results}

\begin{figure}
\begin{center}
\leavevmode
\includegraphics[width=0.75\textwidth]{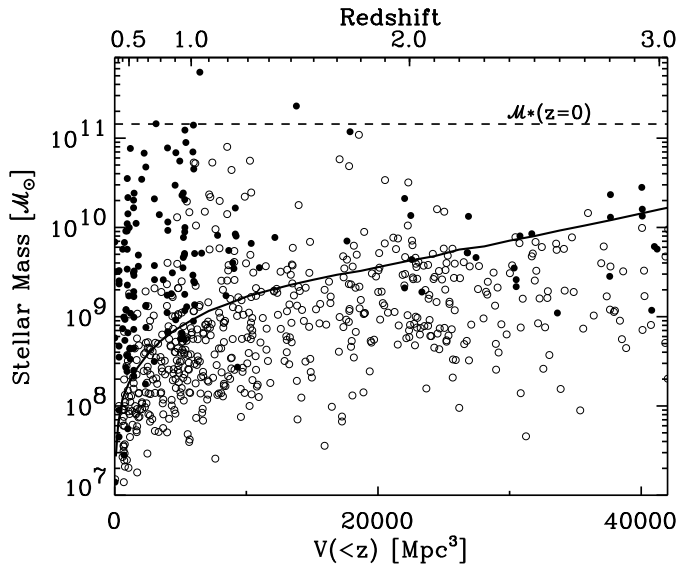}
\end{center}
\caption{Distribution of stellar mass for galaxies in the HDF--N as a
function of co-moving volume. The stellar--mass estimates assume models
with solar metallicity, Salpeter IMF, and single--component
star--formation histories.  Solid symbols denote galaxies with
spectroscopically confirmed redshifts, and open symbols those galaxies
where only photometric redshifts are available.  The
horizontal dashed line indicates the characteristic stellar mass of a
present--day $L^\ast$ galaxy~\cite{col01}.  The solid curve traces the
``mass--limit'' for a maximally old galaxy formed as a burst at $z\sim
\infty$ with passive evolution, and normalized to the flux of the
NICMOS detection limit ($H_\mathrm{AB} \approx 26.5$). }
\label{fig:hdfmass}
\end{figure}

Although at present the stellar--mass estimates for high--redshift
galaxies' have significant uncertainties, these constraints are
interesting nevertheless.  For LBGs with ``$L^\ast$'' UV luminosities
\cite{ste99}, we infer stellar mass estimates of $\sim
10^{10}\;\mathcal{M}_\odot$ or $\sim 1/10$th that of a present--day
$L^\ast$ galaxy \cite{col01}.  Extending this analysis to all galaxies
in the NICMOS HDF--N catalog allows a comparison between the LBG
population and galaxies down to more modest redshifts ($z\mathrel{\hbox{\rlap{\hbox{\lower4pt\hbox{$\sim$}}}\hbox{$>$}}} 0.5$).
In fig.~\ref{fig:hdfmass}, we show the distribution of galaxy stellar
mass in the HDF--N as a function of co-moving volume.  Here, all
stellar masses assume solar metallicity, a Salpeter IMF, and use only
the single--component star--formation histories.  As such, they are
nominally strict lower limits.  Also shown in the figure is a fiducial
curve denoting the minimal detectable stellar mass of a maximally old
galaxy as a function of redshift and the NICMOS detection limit.  Old
galaxies would be detectable with masses \emph{above} this curve.
This, however, does not limit the minimal detectable masses of
galaxies with lower mass--to--light ratios.

\begin{figure}
\begin{center}
\leavevmode
\includegraphics[width=0.75\textwidth]{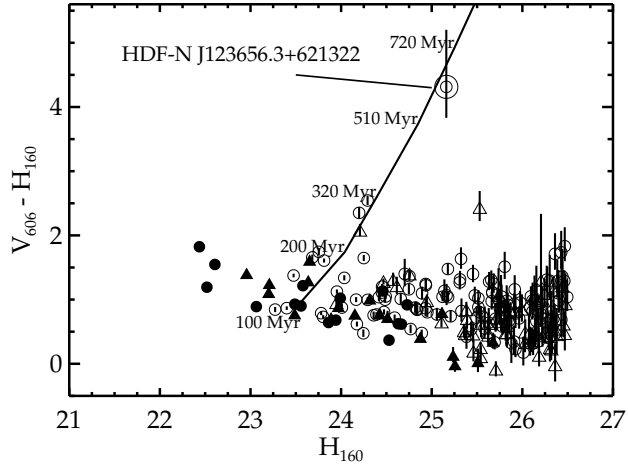}
\end{center}
\caption{Color--magnitude diagram for the HDF--N galaxies with $1.9
\leq z \leq 3.5$.  Solid symbols denote galaxies with
spectroscopically confirmed redshifts, and the open symbols those
galaxies with only photometric redshifts.  The solid curves denotes
the evolution of a $10^{10}\;\mathcal{M}_\odot$ galaxy at $z=2.7$
formed in a $\delta$--function star--formation history. Note that the
``J''--dropout, HDF--N J123656.3+621322 \cite{dic00}, is the only
candidate for an old, red galaxy in the HDF--N in this redshift range.}
\label{fig:vhcmd}
\end{figure}

There are several interesting implications from
fig.~\ref{fig:hdfmass}.  Firstly, the HDF--N exhibits a lack of of
$\mathcal{M}
\mathrel{\hbox{\rlap{\hbox{\lower4pt\hbox{$\sim$}}}\hbox{$>$}}}
\mathcal{M}^\ast(z=0)$ galaxies at
$z\mathrel{\hbox{\rlap{\hbox{\lower4pt\hbox{$\sim$}}}\hbox{$>$}}} 2$.
Such galaxies should be detected (if present) in the deep NICMOS data,
even to $z \sim 3$ (beyond which the NICMOS $H$ band shifts below the
4000~\AA/Balmer break and the stellar mass estimates are less
secure). However, as shown in fig.~\ref{fig:vhcmd}, there are few (if
any) galaxies in this redshift range with $V_{606}-H_{160}$ colors
indicative of a galaxy dominated by old stellar populations.  Thus, it
is unlikely that we are missing them if they were present
(however, see recent results from the HDF--S, e.g., Labb\'e et al.,
this volume).  It is a possibility that we have underestimated their
stellar masses due to the uncertainties described above.   Secondly,
by $z\mathrel{\hbox{\rlap{\hbox{\lower4pt\hbox{$\sim$}}}\hbox{$<$}}}
1.8$, the upper envelope of stellar mass in the HDF--N increases to
include massive, ``$L^\ast$''--sized galaxies. Thus, it seems that the
stellar populations of the progenitors to the massive galaxy
population do not appear to be fully assembled in
$z\mathrel{\hbox{\rlap{\hbox{\lower4pt\hbox{$\sim$}}}\hbox{$>$}}} 2$
progenitors.  This in turn suggests that more star--formation or
merging (or both) are required for
$z\mathrel{\hbox{\rlap{\hbox{\lower4pt\hbox{$\sim$}}}\hbox{$<$}}} 2$
to construct the large--galaxy population observed at
$z\mathrel{\hbox{\rlap{\hbox{\lower4pt\hbox{$\sim$}}}\hbox{$<$}}} 1$
and at the present--epoch.

We wish to thank the conference organizers for arranging such a
stimulating meeting in a beautiful setting. Support for this work was
provided by NASA through grant GO--07817.01-96A.

%


\begin{thebibliography}{8.}
\addcontentsline{toc}{section}{References}

\bibitem{dic01}
M. Dickinson, et al.: \emph{in preparation}

\bibitem{pap01}
C. Papovich, M. Dickinson, \& H. C. Ferguson: ApJ \textbf{559}, 620 (2001)

\bibitem{bru93}	
G. A. Bruzual, \& S. Charlot: ApJ \textbf{405}, 538 (1993)

\bibitem{bru01} 
G. A. Bruzual: Ap\&SS Sup.\ \textbf{227}, 221 (2001)

\bibitem{col01}
S. Cole, et al.: MNRAS \textbf{326}, 255 (2001)

\bibitem{pet01}
M. Pettini,  et al.: ApJ \textbf{554}, 981 (2001)

\bibitem{dic00}
M. Dickinson, et al.: ApJ \textbf{531}, 624, (2000)

\bibitem{ste99} C. C. Steidel, et al.: ApJ \textbf{519}, 1 (1999)




\end{thebibliography}
\end{document}